\documentclass[conference， letterpaper， 10pt]{IEEEtran}%a4paper,10pt
 \usepackage{amsfonts}
 \IEEEoverridecommandlockouts
\usepackage{amsmath}
\usepackage[letterpaper, top=1in, bottom=1.1in, left=0.75in, right=0.75in]{geometry}
\usepackage{epsfig}
\usepackage{graphicx}
\usepackage{psfig}
\usepackage{subcaption}
\usepackage{epsf}
\usepackage{epstopdf}
\usepackage{bbm}
\usepackage{colortbl}
\usepackage[most]{tcolorbox}
\usepackage{tikz}
\usetikzlibrary{arrows.meta,shapes,decorations,automata,backgrounds,petri,topaths,calc,mindmap,trees,positioning,chains,arrows}
\usepackage{relsize}
\usepackage{pgfplots}
\pgfplotsset{compat=1.14}
 \usepackage{algorithm2e}
\usepackage{algpseudocode}
\usepackage{amssymb}
\usepackage{float}
\usepackage{multirow}
\usepackage{pbox}

\usepackage{amsmath}
\allowdisplaybreaks[4]
\usepackage{booktabs}
\usepackage{fancyhdr}

\usepackage{epstopdf, cite,color}
\usepackage{amsthm}
\usepackage{amssymb}
\newcommand{\bc}{\begin{center}}
	\newcommand{\ec}{\end{center}}
\newcommand{\be}{\begin{equation}}
\newcommand{\ee}{\end{equation}}
\newcommand{\bea}{\begin{eqnarray}}
\newcommand{\eea}{\end{eqnarray}}

\title{LLM‑Assisted Alpha‑Fairness for 6 GHz Wi‑Fi/NR‑U Coexistence: An Agentic Orchestrator for Throughput, Energy, and SLA}

\begin{document}

\author{ Qun Wang\textsuperscript{\S},Yingzhou Lu \textsuperscript{{\dag}}, Guiran Liu\textsuperscript{\S}, Binrong Zhu\textsuperscript{\S}, and Yang Liu\textsuperscript{\S}\\
    
	 \textsuperscript{\S}Department of Computer Science, San Francisco State University, San Francisco, CA, 94132\\
\textsuperscript{{\dag}}School of Medicine, Stanford University, USA\\
    Emails:  
       claudqunwang@ieee.org, lyz66@stanford.edu, \{gliu, bzhu2, yliu68\}@sfsu.edu,
	
    }

\maketitle

\begin{abstract}
  Unlicensed 6\,GHz is becoming a primary workhorse for high-capacity access, with Wi-Fi and 5G NR-U competing for the same channels under listen-before-talk (LBT) rules. Operating in this regime requires decisions that jointly trade throughput, energy, and service-level objectives while remaining safe and auditable. We present an agentic controller that separates {policy} from {execution}. At the start of each scheduling epoch the agent summarizes telemetry (per-channel busy and baseline LBT failure; per-user CQI, backlog, latency, battery, priority, and power mode) and invokes a large language model (LLM) to propose a small set of interpretable knobs: a fairness index $\alpha$, per-channel duty-cycle caps for Wi-Fi/NR-U, and class weights. A deterministic optimizer then enforces feasibility and computes an $\alpha$-fair allocation that internalizes LBT losses and energy cost; malformed or unsafe policies are clamped and fall back to a rule baseline. In a 6\,GHz simulator with two 160\,MHz channels and mixed Wi-Fi/NR-U users, LLM-assisted policies consistently improve energy efficiency while keeping throughput competitive with a strong rule baseline. One LLM lowers total energy by $35.3\%$ at modest throughput loss, and another attains the best overall trade-off, finishing with higher total bits ($+3.5\%$) and higher bits/J ($+12.2\%$) than the baseline. We release code, per-epoch logs, and plotting utilities to reproduce all figures and numbers, illustrating how transparent, policy-level LLM guidance can safely improve wireless coexistence.
\end{abstract}

\section{Introduction}

Unlicensed 6\,GHz spectrum is rapidly becoming a workhorse for high-capacity wireless access, with Wi-Fi 6E/7 and 5G NR-U expected to coexist in the same channels \cite{qunss}. The resulting environment is shaped by LBT rules, bursty traffic, heterogeneous device batteries and priorities, and tight latency targets for interactive and safety-critical applications. Conventional schedulers are typically crafted around a fixed objective, such as sum-rate maximization or proportional fairness, and rely on static heuristics to avoid collisions. While effective in specific regimes, these designs struggle to adapt when operating conditions drift, and they provide limited knobs to navigate the three-way tension between throughput, energy, and service-level objectives \cite{qunirsss}.

 State-of-the-art solutions for managing spectrum coexistence between technologies like Wi-Fi and NR-U predominantly fall into two categories: static rule-based schedulers and complex learning-based systems like deep reinforcement learning (DRL) \cite{ruless}\cite{drlss1}. Rule-based schedulers are reliable and predictable but are fundamentally brittle; they are designed with fixed priorities, such as maximizing overall throughput, and struggle to dynamically adapt when network conditions and objectives change \cite{ruless}. For example, they cannot easily pivot to prioritize energy efficiency or meet a specific user's urgent latency demand when the network becomes congested . On the other hand, while DRL agents can learn sophisticated policies to manage these trade-offs, they often operate as "black boxes," making their decisions difficult to interpret, audit, or trust in a production network \cite{drlss1} . Furthermore, DRL systems require extensive, often impractical, online training cycles to converge on an effective policy. 
 
 Our paper solves this critical gap by introducing an agentic controller that separates high-level reasoning from low-level execution. It leverages a LLM to interpret the complex, multi-faceted network state and propose a simple, human-understandable set of policy "knobs"—such as the fairness level  and duty-cycle caps \cite{ylu1} \cite{ylu2}. This approach avoids the rigidity of fixed rules and the opacity of DRL, providing a solution that is simultaneously adaptive, auditable, and capable of intelligently balancing the conflicting objectives of throughput, energy efficiency, and service level agreements (SLA) on a dynamic, per-epoch basis.

 At the start of each scheduling epoch, the agent encodes the observable state—per-channel busy levels and baseline LBT failure probabilities for both Wi-Fi and NR-U, together with per-user CQI, backlog, latency target, battery state, task priority, and power mode—into a compact JSON summary. A high-level policy then chooses a small set of interpretable knobs: a fairness index, per-channel duty-cycle caps for each stack, and priority weights across traffic classes. This policy can be instantiated either as a deterministic rule or as a LLM prompted to reason about energy-aware, latency-aware spectrum sharing.
% The proposed knobs are sanitized by range checks and headroom constraints derived from sensed channel occupancy, after which a deterministic optimizer enforces feasibility and computes an $\alpha$-fair allocation that internalizes LBT loss and energy cost. This separation keeps the LLM’s role transparent and auditable—limited to high-level trade-offs—while guaranteeing that the executed schedule satisfies hard constraints.

% The optimizer follows a two-stage pattern that has proved robust for coexistence. First, it assigns each user to a single channel by maximizing a probe-time utility density that weighs post-LBT goodput against energy and deadline pressure. Second, within each channel and stack, it grants minimal duties to urgent flows and distributes the remaining budget via a weighted $\alpha$-fair split. Losses are realized from the final aggregate duties, and energy is computed from a mode-dependent per-bit cost. The system logs per-epoch served bits, energy, SLA hits, and the chosen $\alpha$, enabling end-to-end reproducibility and ablation.

We evaluate the agent in a 6\,GHz simulator with two 160\,MHz channels and mixed Wi-Fi/NR-U populations. Across moderate and high offered loads, LLM-assisted policies consistently reduce cumulative energy and improve energy efficiency (bits/J) relative to a strong rule baseline, while maintaining competitive or superior cumulative throughput. In a representative scenario, one LLM variant lowers total energy by more than a third while the other achieves the best overall trade-off, finishing with higher total bits and the highest bits/J. 
% These gains arise from headroom-aware duty caps and fairness choices closer to proportional fairness, which steer the system away from collision-dominated operating points without starving progress.

% This paper contributes a coherent formulation of LLM-assisted spectrum orchestration for Wi-Fi/NR-U coexistence, a safety-first interface that constrains free-form model outputs to verifiable controls, and a lightweight optimizer that realizes $\alpha$-fair allocations under LBT and energy models. We provide an anonymized artifact with code, per-epoch logs, and plotting utilities to reproduce all figures and numbers reported here. The results suggest that transparent, policy-level LLM guidance, when coupled with a deterministic executor, can materially improve energy-throughput trade-offs in unlicensed-band coexistence while preserving operational safety and reproducibility.

The remainder of this paper is structured as follows. In Section II, we detail the system model and formulate the multi-objective resource allocation problem. Section III presents the design of the LLM-assisted agentic controller. We describe our simulation environment and the baseline rule-based scheduler used for comparison in Section IV. In Section V, we present a comprehensive evaluation of our proposed system, analyzing its performance across key metrics of network throughput, energy efficiency, and SLA satisfaction against the baseline. Finally, we conclude the paper in Section VI.

\section{System Model and Problem Formulation}
\label{sec:system-model}

\subsection{System Model}

We consider a 6\,GHz unlicensed band shared by Wi‑Fi and 5G NR‑U. Time is slotted into fixed scheduling epochs of length $\Delta$ seconds (default $\Delta=0.1$), and all control decisions are recomputed once per epoch $t=1,2,\ldots$ The system comprises a finite set of channels $\mathcal{C}$ (two 160\,MHz channels in the default configuration) and two coexisting technology stacks $\mathcal{T}=\{\text{Wi‑Fi},\text{NR‑U}\}$. Each channel $c\in\mathcal{C}$ has bandwidth $B_c$ and is characterized by exogenous LBT conditions that vary slowly over time. Specifically, for each stack $t\in\mathcal{T}$ and channel $c$, we track a sensed busy fraction $b_{t,c}(t)\in[0,1]$ and a baseline LBT failure probability $f_{t,c}(t)\in[0,1]$, both subjected to small random jitter between epochs to emulate environmental dynamics. These quantities are taken as inputs to the scheduler and are observable before each decision. 

Users are partitioned by stack, $\mathcal{U}=\mathcal{U}_{\text{Wi‑Fi}}\cup\mathcal{U}_{\text{NR‑U}}$. A user $i\in\mathcal{U}$ is described at epoch $t$ by a channel quality indicator $q_i(t)\in\{0,\ldots,15\}$, a battery state $B_i(t)\in[0,1]$, a queue backlog $Q_i(t)$ in bits, a latency target $D_i$ in milliseconds, a task priority class $k_i\in\{\text{emergency, high, normal, bulk}\}$, and a discrete power mode $p_i\in\{\text{low, med, high}\}$. New traffic arrivals $A_i(t)$ are injected into the queue each epoch according to a truncated Gaussian process with a configurable mean; the queue dynamics obey
\begin{equation}
Q_i(t{+}1)\;=\;\max\bigl\{\,0,\; Q_i(t)+A_i(t)-S_i(t)\,\bigr\},
\label{eq:queue}
\end{equation}
where $S_i(t)$ denotes the number of bits served to user $i$ during epoch $t$ \cite{qunlpy}. The simulation exposes these elements as first-class state variables that are evolved by a lightweight environment model prior to each scheduling step. 

Physical-layer throughput is modeled via a CQI-to-spectral-efficiency mapping. For user $i$ on channel $c$, the spectral efficiency (bits/s/Hz) is
\begin{equation}
s_{i,c}(t)\;=\;\operatorname{SE}\!\bigl(q_i(t)\bigr)\cdot \eta_{p_i},
\label{eq:se}
\end{equation}
where $\operatorname{SE}(\cdot)$ is a standard 16-level table and $\eta_{p_i}$ scales the efficiency under the user’s power mode. If $\tau_{i,c}(t)\in[0,1]$ is the airtime fraction allotted to $i$ on $c$ during epoch $t$, then the pre‑LBT raw rate is
\begin{equation}
r_{i,c}(t)\;=\; s_{i,c}(t)\, B_c\, \tau_{i,c}(t).
\label{eq:raw-rate}
\end{equation}
Shared-channel contention and regulatory LBT constraints induce a stack–channel loss that we capture with a smooth proxy. Let $\tau_{t,c}(t)=\sum_{i\in\mathcal{U}_t}\tau_{i,c}(t)$ be the aggregate airtime of stack $t$ on channel $c$ in epoch $t$. The loss fraction applied uniformly to all users of $(t,c)$ is
\begin{equation}
\begin{split}
\ell_{t,c}(t)\;=\;\min\!\Bigl\{\,0.95,\; f_{t,c}(t)\;+\;0.6\,\tau_{t,c}(t)\,b_{t,c}(t)\;\\
+\;0.2\,\bigl(\tau_{t,c}(t)+b_{t,c}(t)-1\bigr)_{+}\Bigr\},
\end{split}
\label{eq:lbt}
\end{equation}
with $(x)_{+}=\max\{x,0\}$. The post‑LBT goodput for user $i$ on $c$ is then
\begin{equation}
g_{i,c}(t)\;=\; r_{i,c}(t)\,\bigl(1-\ell_{t_i,c}(t)\bigr),
\end{equation}
and

\begin{equation}
S_i(t)\;=\; \Delta\sum_{c\in\mathcal{C}} g_{i,c}(t).
\label{eq:goodput}
\end{equation}
Energy is modeled through a per‑bit cost that depends on the power mode and the achieved spectral efficiency. Writing $P(p_i)$ for the mode‑dependent transmit power and $e_{i,c}(t)=P(p_i)/s_{i,c}(t)$ for joules per bit, the per‑epoch energy is
\begin{equation}
E_i(t)\;=\;\sum_{c\in\mathcal{C}} e_{i,c}(t)\,S_i(t).
\label{eq:energy}
\end{equation}
The above link, loss, and energy models mirror the implementation used by our simulator, where \eqref{eq:lbt} is realized as a differentiable proxy to capture LBT‑driven collisions/backoff at the stack–channel granularity. 

Service-level latency is expressed as an instantaneous rate requirement. Given backlog $Q_i(t)$ and latency target $D_i$, the minimum rate that avoids deadline slippage within epoch $t$ is
\begin{equation}
\rho_i(t)\;=\;\min\!\Bigl\{\tfrac{Q_i(t)}{\Delta},\; \tfrac{Q_i(t)}{D_i/1000}\Bigr\},
\label{eq:sla}
\end{equation}
and an SLA hit is registered whenever $\sum_{c} g_{i,c}(t)\ge \rho_i(t)$. The policy layer that precedes optimization provides three high‑level knobs per epoch: a fairness index $\alpha\in\{0,1,2\}$, per‑channel duty‑cycle caps $u^{\text{Wi‑Fi}}_{c},u^{\text{NR‑U}}_{c}\in[0,1]$, and priority weights $\{w_k\}_{k\in\{\text{emergency,high,normal,bulk}\}}$. Caps are constrained by sensed load through a headroom rule of the form $u^t_c\le 1-\gamma\,b_{t,c}(t)$ with $\gamma\in(0,1)$, ensuring that aggregate scheduled airtime remains feasible under exogenous activity. These quantities are produced either by a deterministic rule or by a LLM from a compact JSON state summary, and are clamped to safe ranges before optimization. 

\subsection{Problem Formulation}

Let $x_{i,c}(t)\in\{0,1\}$ indicate whether user $i$ is scheduled on channel $c$ during epoch $t$. Each user is bound to at most one channel per epoch, $\sum_{c} x_{i,c}(t)\le 1$, and receives a nonnegative airtime $\tau_{i,c}(t)\in[0,1]$ that is consistent with the assignment via $\tau_{i,c}(t)\le x_{i,c}(t)$. Per‑stack duty caps provided by the policy enforce
\begin{equation}
\sum_{i\in\mathcal{U}_{\text{Wi‑Fi}}}\tau_{i,c}(t)\ \le\ u^{\text{Wi‑Fi}}_{c},
\end{equation}
\begin{equation}
\sum_{i\in\mathcal{U}_{\text{NR‑U}}}\tau_{i,c}(t)\ \le\ u^{\text{NR‑U}}_{c},
\quad \forall c\in\mathcal{C},
\label{eq:cap-constraints}
\end{equation}
and interact with LBT through \eqref{eq:lbt}–\eqref{eq:goodput}. The per‑user post‑LBT rate in epoch $t$ is $x_i(t)=\sum_{c} g_{i,c}(t)$, and the per‑epoch energy is $E_i(t)$ in \eqref{eq:energy}. A standard weighted $\alpha$‑fair utility is adopted \cite{qunfairness},
\begin{equation}
U_{\alpha}(x)\;=\;
\begin{cases}
\log x, & \alpha=1,\\
\dfrac{x^{\,1-\alpha}}{1-\alpha}, & \alpha\neq 1,
\end{cases}
\qquad x>0,
\label{eq:alpha}
\end{equation}
and each user $i$ is endowed with a base weight $\theta_i=\dfrac{w_{k_i}}{0.5+\beta(B_i)}$ that increases with priority and penalizes low battery via a monotone function $\beta(\cdot)$. 

The single‑epoch allocation problem can then be stated as
\begin{align}
\max_{\{x_{i,c},\,\tau_{i,c}\}}\quad
& \sum_{i\in\mathcal{U}} \theta_i\,U_{\alpha}\!\left(\sum_{c\in\mathcal{C}} g_{i,c}(t)\right)\;-\;\lambda\sum_{i\in\mathcal{U}} E_i(t)
\label{eq:master-obj}\\
\text{s.t.}\quad
& \text{link and LBT coupling in \eqref{eq:raw-rate}–\eqref{eq:goodput}},\nonumber\\
& \text{caps in \eqref{eq:cap-constraints}},\nonumber\\
&\tau_{i,c}(t)\in[0,1],\ x_{i,c}(t)\in\{0,1\},\nonumber\\
&\sum_{c} x_{i,c}(t)\le 1.\nonumber
\end{align}
Here $\lambda\ge 0$ trades off energy against throughput in the objective; in our implementation the energy term is effectively captured by the interaction of \eqref{eq:energy} with the weights $\theta_i$ and per‑duty utility densities used during channel selection, while the reported $\alpha$‑fair utility remains the primary figure of merit. Hard per‑epoch latency guarantees may be included as $\sum_{c} g_{i,c}(t)\ge \rho_i(t)$ for selected flows, but we favor a pragmatic approach in which minimum “urgent” grants are applied procedurally before the proportional $\alpha$‑fair split on the remaining budget—an approach that preserves tractability while honoring latency‐sensitive traffic. 

Problem~\eqref{eq:master-obj} is solved independently at each epoch using policy‑provided knobs $(\alpha,\{u^t_c\},\{w_k\})$, yielding the airtime variables $\{\tau_{i,c}(t)\}$, post‑LBT rates $\{g_{i,c}(t)\}$, and per‑epoch metrics (throughput, energy, and SLA hit rate) that feed back into the next epoch via \eqref{eq:queue}. This separation between a high‑level, possibly LLM‑driven policy and a verifiable optimizer enables safe orchestration: policy outputs are clamped to feasible ranges, and the optimizer enforces hard constraints at execution time.

\section{Method: Agent Architecture}
\label{sec:method}

This work adopts an agentic architecture that separates high-level spectrum \emph{policy} from low-level \emph{optimization}. The policy layer proposes a fairness index $\alpha$, per-channel duty-cycle caps for Wi‑Fi and NR‑U, and task-class priority weights from observable telemetry; the optimizer then enforces feasibility and computes an $\alpha$‑fair allocation while accounting for LBT losses, energy, and latency. The separation ensures that the intelligence responsible for strategic trade-offs is modular and replaceable (rule-based or LLM-driven), whereas the executor is verifiable and deterministic. The implementation follows this design in a single-epoch loop with environmental evolution between epochs. 

\subsection{Telemetry encoding and policy interface}

At the beginning of each epoch of length $\Delta$, the environment exposes a compact JSON state containing channel descriptors (bandwidth $B_c$, sensed busy $b_{t,c}$, and baseline LBT failure $f_{t,c}$ for each stack $t$ on channel $c$) and user descriptors (CQI $q_i$, battery level $B_i$, backlog $Q_i$, latency target $D_i$, task priority $k_i$, and power mode $p_i$). The function \texttt{build\_state\_json} produces this state and adds lightweight hints such as candidate $\alpha$ values. A policy consumes this JSON and returns three objects: an $\alpha\in\{0,1,2\}$; duty caps $\{u^{\mathrm{Wi\text{-}Fi}}_c,u^{\mathrm{NR\text{-}U}}_c\}$; and priority weights $\{w_k\}$ for classes \texttt{emergency/high/normal/bulk}. The project ships two variants of the policy interface: a rule baseline that biases caps toward the busier stack while reserving headroom, and an LLM policy that emits a JSON policy either via a chat completion in JSON mode or via a Responses API call constrained by a JSON Schema; in both cases the returned values are coerced to safe ranges prior to optimization. 

\subsection{Safety, feasibility, and fallbacks}
 The fairness index is restricted to $\{0,1,2\}$; duty caps are clipped to $[0,1]$ and to a busy-aware headroom $u^t_c \le 1-\gamma\,b_{t,c}$ (with $\gamma\in(0,1)$ implemented as $0.5$ in the code); and priority weights are confined to a bounded interval $[0.1,10]$. If the LLM call fails or returns malformed JSON, the system falls back to the deterministic rule policy, ensuring that every epoch yields a feasible control. Optional textual rationales from the LLM are preserved for auditability but do not affect the optimizer.

\subsection{Epoch solver: two-stage optimization}

Given the policy knobs, the optimizer solves the epoch using a two-stage scheme tailored to coexistence with shared LBT losses.

The first stage assigns each user to one channel by maximizing a \emph{utility density} computed under a small probe airtime $\tau_0$, which approximates the value per unit of duty cycle while internalizing energy and latency costs. For user $i$ on channel $c$, the pre-loss rate is $r_{i,c}=s_{i,c} B_c \tau_0$, where $s_{i,c}$ is the CQI- and power-mode–dependent spectral efficiency. The stack-channel loss is modeled by a smooth proxy:
\begin{equation}
\begin{split}
\ell_{t,c}=\min\{0.95, f_{t,c} + 0.6\,\tau_{t,c}\,b_{t,c}\\ 
+ 0.2\,(\tau_{t,c}+b_{t,c}-1)_+\},
\label{eq:lbt-proxy-method}
\end{split}
\end{equation}
with $\tau_{t,c}$ being the aggregate duty of stack $t$ on channel $c$. The probe goodput is $g_{i,c}=r_{i,c}(1-\ell_{t_i,c})$, and the corresponding energy consumed during the probe is $E_{i,c}^{\text{probe}}$. The assignment score reflects a direct trade-off between this goodput and the energy cost, defined as:
\begin{equation}
\begin{split}
\Phi_{i,c}\;=\;\frac{1}{\tau_0}\left( \underbrace{w_{k_i} \cdot \frac{g_{i,c}}{10^6} \cdot w_{\mathrm{lat}}(D_i)}_{\text{Reward}} - \underbrace{\beta(B_i) \cdot E_{i,c}^{\text{probe}}}_{\text{Cost}} \right),
\end{split}
\label{eq:utility-density-method}
\end{equation}
where $w_{k_i}$ is the user's priority weight, $w_{\mathrm{lat}}(D_i)$ is a multiplier that increases the reward for latency-sensitive users (e.g., those with $D_i \le 50$~ms), and $\beta(B_i)$ is a penalty factor that increases as battery level $B_i$ decreases. The channel with the maximal score $\Phi_{i,c}$ is chosen for user $i$. This stage has complexity $O(|\mathcal{U}|\,|\mathcal{C}|)$ and captures the primary cross-channel trade-offs.

The second stage performs within-channel allocation under the duty caps provided by the policy. For each channel and stack, the available duty budget $\le u^t_c$ is split in two passes. A first pass grants \emph{urgent minimums} to latency-critical or high-priority users, allocating the minimal duty required to meet their instantaneous rate requirement $\rho_i=\min\{Q_i/\Delta,\,Q_i/(D_i/1000)\}$. A second pass distributes the residual budget via a weighted $\alpha$‑fair rule, where user $i$ receives a portion of the airtime proportional to:
\begin{equation}
\omega_i\;=\;\left(\frac{w_{k_i}}{0.5+\beta(B_i)}\right)\cdot\bigl(\mathrm{served}_i+\varepsilon\bigr)^{-\alpha}.
\label{eq:alpha-weights-method}
\end{equation}
Here, $\mathrm{served}_i$ is the service a user has already received within the epoch (from the urgent grant) and $\varepsilon>0$ is a small constant to ensure stability. After all duties are provisionally assigned, the final aggregate duties are used to recompute the stack-channel losses $\ell_{t,c}$, from which the final per-user goodput and energy consumption are determined.

\subsection{SLA evaluation, queue update, and logging}

Following allocation, the achieved per-user rate $\sum_c g_{i,c}$ is compared against $\rho_i$ to determine SLA hits in the current epoch. The served bits are subtracted from backlogs to update $Q_i$ for the next epoch, ensuring tight coupling between control and traffic dynamics. In multi-epoch mode, the driver \texttt{run\_multi\_epoch} evolves channels and baselines with small Gaussian jitters, injects arrivals with a configurable mean, repeatedly invokes the epoch solver.

% \subsection{Complexity and extensibility}

% The dominant cost arises in stage one, which is linear in the product of users and channels due to the evaluation of \eqref{eq:utility-density-method}; stage two is linear in the number of assigned users per channel due to proportional allocation after urgent grants. The modular boundary between policy and optimizer makes several extensions straightforward: integer programming to model RU/RB discreteness while keeping the same policy interface; smoothing or hysteresis on policy knobs to stabilize control; or offline distillation of $(\text{state},\text{policy},\text{outcome})$ triplets into a lightweight policy network with occasional LLM intervention for unfamiliar regimes. The codebase includes two policy-call implementations—JSON-mode chat completions and JSON-Schema–constrained responses—so that experiments can switch between rule and LLM orchestration without touching the optimizer. 

\subsection{LLM-Assisted Decision Making}

At the beginning of each epoch, the agent summarizes the observable telemetry into a compact JSON state that includes per-channel descriptors (bandwidth, sensed busy fractions, and baseline LBT failure rates for each stack) and per-user descriptors (CQI, battery level, backlog, latency target, task priority, and power mode). This serialization, produced by \texttt{build\_state\_json}, is passed to a large language model that proposes high-level control knobs: a fairness index $\alpha\in\{0,1,2\}$, per-channel duty-cycle caps for Wi-Fi and NR-U, and task-class priority weights. We implement two invocation modes to ensure robustness: a chat-completions call constrained to JSON output, and a Responses-API call that enforces a JSON Schema with strict types and bounds. In both modes the model is prompted as a spectrum policy orchestrator and asked to trade off latency, energy, and fairness while avoiding per-user micromanagement. The LLM may return brief rationales, which are logged for audit but are not used downstream in optimization. 

The raw policy is never executed directly. Instead, \texttt{coerce\_policy\_from\_llm} enforces feasibility and safety by clamping $\alpha$ to $\{0,1,2\}$, projecting duty caps to $[0,1]$ and to a busy-aware headroom of the form $u^{t}_{c} \le 1 - \gamma\, b_{t,c}$ (with $\gamma\in(0,1)$), and restricting priority weights to a bounded interval that prevents extreme allocations. Any parsing failure, schema violation, or out-of-range proposal triggers a deterministic rule fallback that biases caps toward the empirically busier stack while reserving headroom, guaranteeing that every epoch yields a valid policy even under LLM faults. 

Given the sanitized knobs $(\alpha,\{u^{t}_{c}\}_{c,t},\{w_{k}\}_{k})$, the optimizer solves the epoch in two stages. First, it assigns each user to a single channel by maximizing a probe-time utility density that internalizes post-LBT goodput, energy cost, deadline pressure, and priority/battery weights. Second, within each channel and stack, it grants minimal duties to satisfy urgent latency targets and then allocates the remaining budget according to a weighted $\alpha$-fair rule. After duties are finalized, stack–channel LBT losses are realized and per-user goodputs and energies are computed. The selected $\alpha$ and per-epoch metrics (throughput, energy, SLA hit rate) are recorded, and the agent advances to the next epoch with updated queues. This integration makes the LLM responsible only for transparent, high-level decisions, while a verifiable executor enforces hard constraints at run time.

The core of the interaction between the agent and the LLM is a structured JSON object that serves as the complete state representation provided at the start of each scheduling epoch. 
The JSON object is organized into two primary keys: \texttt{channels} and \texttt{users}.
\textbf{\texttt{channels}}: An array of objects detailing the physical state and contention level of each frequency channel. For each channel, we provide its bandwidth (\texttt{bw\_mhz}), the measured busy-time contributed by Wi-Fi and NR-U (\texttt{busy\_wifi}, \texttt{busy\_nru}), and the baseline LBT failure probability.
\textbf{\texttt{users}}: An array of objects representing the state of each active user. For each user, the prompt specifies their technology (\texttt{tech}), channel quality indicator (\texttt{cqi}), data backlog in bits (\texttt{backlog\_bits}), remaining time to meet their SLA deadline (\texttt{deadline\_s}), battery percentage (\texttt{battery\_pct}), and assigned service priority class (\texttt{priority}).

% \begin{figure}[h]
% \begin{lstlisting}[language=json, caption={Example JSON prompt representing the network state for the LLM.}, label={lst:json_prompt}, basicstyle=\footnotesize\ttfamily, breaklines=true]
% {
%   "channels": [
%     {
%       "id": 0,
%       "bw_mhz": 160,
%       "busy_wifi": 0.25,
%       "busy_nru": 0.45,
%       "lbt_fail_prob": {"wifi": 0.15, "nru": 0.20}
%     }
%   ],
%   "users": [
%     {
%       "id": 0,
%       "tech": "wifi",
%       "cqi": 8,
%       "backlog_bits": 5000000,
%       "deadline_s": 0.08,
%       "battery_pct": 75,
%       "priority": "high"
%     },
%     {
%       "id": 1,
%       "tech": "nru",
%       "cqi": 10,
%       "backlog_bits": 800000,
%       "deadline_s": 0.5,
%       "battery_pct": 30,
%       "priority": "normal"
%     }
%   ]
% }
% \end{lstlisting}
% \end{figure}

\section{Experimental Setup and Results}
\label{sec:exp}

\subsection{Experimental Setup}

We evaluate the agent in a simulated 6\,GHz unlicensed band with two 160\,MHz channels shared by Wi‑Fi and NR‑U. Each experiment spans $T{=}100$ scheduling epochs of length $\Delta{=}0.1$\,s (total horizon 10\,s) \cite{6ghz1}. The default user population includes 16 Wi‑Fi and 12 NR‑U stations with heterogeneous channel quality indicators (CQI), battery levels, queue backlogs, latency targets, task priorities, and power modes. Channel descriptors include sensed busy fractions and baseline LBT failure probabilities for both stacks, all subject to small Gaussian jitter between epochs to emulate environmental dynamics. Traffic arrivals are injected every epoch as truncated Gaussians, and queue evolution follows the standard Lindley recursion. The simulator exposes these elements as first‑class state variables ($\mathsf{User}$, $\mathsf{Channel}$, $\mathsf{Env}$) and advances them via \texttt{step\_env} before each decision; the single‑epoch solver is called from \texttt{one\_epoch\_allocate}, and multi‑epoch orchestration from \texttt{run\_multi\_epoch}. The code also logs per‑epoch throughput (served bits), energy (J), SLA hit rate, and the fairness index $\alpha$.  The code is available at: https://github.com/claudwq/LLM-Assisted-Alpha-Fairness-for-6-GHz-Wi-Fi-NR-U-Coexistence.git

The policy layer is either rule‑based or LLM‑driven. The rule baseline computes per‑channel duty‑cycle caps for the two stacks, reserving headroom as a function of sensed busy, and then chooses the best‑throughput $\alpha\!\in\!\{0,1,2\}$ each epoch (“benevolent” baseline). The LLM policy sees the same telemetry as a compact JSON, selects a single $\alpha$, per‑channel caps, and class weights, and is then clamped by \texttt{coerce\_policy\_from\_llm} for safety before the optimizer is invoked. The optimizer itself is deterministic: it assigns a single channel per user via a utility‑density score that internalizes post‑LBT goodput, energy cost, and deadline pressure, and then performs a within‑channel weighted $\alpha$‑fair split under the cap constraints, realizing LBT loss and energy using the final aggregate duties. The complete path and data schema are implemented in \texttt{llm\_spectrum\_agent\_fairness.py} and its LLM interface variant. 

We compare three methods: \textbf{RuleBased} (no LLM, benevolent $\alpha$), \textbf{GPT4o‑Mini} (LLM‑assisted policy), and \textbf{GPT5‑Mini} (LLM‑assisted policy) \cite{gpt45}. 
All experiments use seed 2025 and the default arrival and jitter settings in the code unless otherwise noted. 

\section{Results Analysis}
\label{sec:results}

We report results for three policies—\emph{RuleBased} (benevolent $\alpha\!\in\!\{0,1,2\}$ chosen per epoch), \emph{GPT4o-Mini}, and \emph{GPT5-Mini}—over $T{=}100$ epochs with $\Delta{=}0.1$\,s. Metrics are computed directly from the simulator logs: per-epoch served bits, energy (J), and SLA hit rate; we plot cumulative throughput (Gb), cumulative energy (J), and cumulative energy efficiency (bits/J). 

\subsection{Moderate Offered Load (40\,Mb/s)}
Under the moderate arrival rate, cumulative throughput rises rapidly during the first 20--30 epochs as the scheduler drains initial backlogs, then transitions to an arrival-limited regime in which curves flatten. In this setting, \emph{GPT5-Mini} ultimately surpasses the baseline in total delivered bits while spending less energy, whereas \emph{GPT4o-Mini} attains the lowest energy but at the cost of reduced throughput. The cumulative energy plots show that \emph{RuleBased} expends the most energy across the horizon; this aligns with its tendency to choose $\alpha{=}0$ and to push higher duties into collision-prone regions. The cumulative energy-efficiency trajectories confirm the advantage of LLM-assisted control: both LLM policies maintain higher bits/J than the rule baseline for most of the horizon, with \emph{GPT5-Mini} finishing highest. Intuitively, the LLMs propose headroom-aware caps and a fairness regime closer to $\alpha{=}1$, which lowers collision losses without starving progress, yielding better energy efficiency at comparable or higher cumulative bits.

% \begin{figure}[t]
%   \centering
%   \includegraphics[width=\linewidth]{40MB-throughput_all.png}
%   \vspace{-0.6em}
%   \caption{Moderate load (40\,Mb/s): cumulative throughput over 100 epochs.}
%   \label{fig:40-throughput}
% \end{figure}

% \begin{figure}[t]
%   \centering
%   \includegraphics[width=\linewidth]{40MB-energy_all.png}
%   \vspace{-0.6em}
%   \caption{Moderate load (40\,Mb/s): cumulative energy over 100 epochs.}
%   \label{fig:40-energy}
% \end{figure}

% \begin{figure}[t]
%   \centering
%   \includegraphics[width=\linewidth]{40MB-EE.png}
%   \vspace{-0.6em}
%   \caption{Moderate load (40\,Mb/s): cumulative energy efficiency (bits/J).}
%   \label{fig:40-ee}
% \end{figure}

\begin{figure*}[t]
  \centering
  \begin{subfigure}{0.3\linewidth}
    \includegraphics[width=\linewidth]{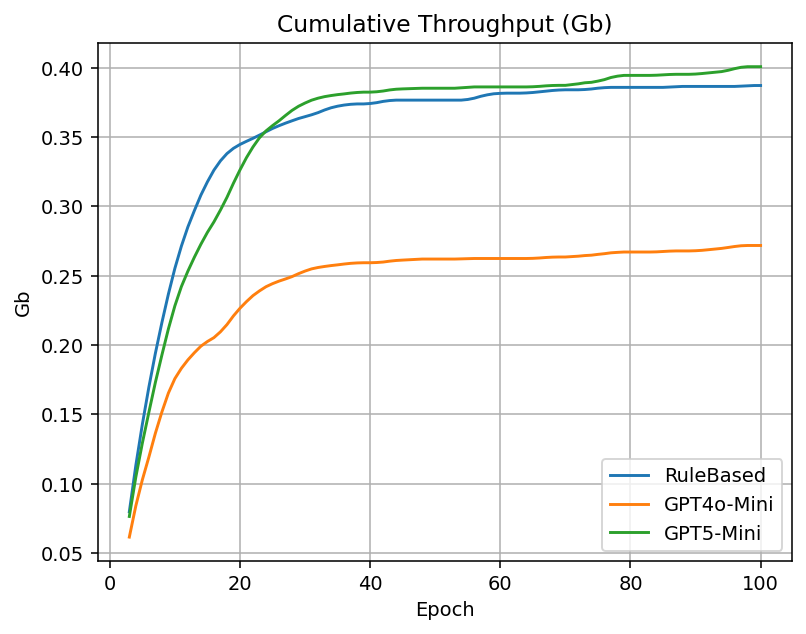}
    \caption{Throughput}
  \end{subfigure}\hfill
  \begin{subfigure}{0.3\linewidth}
    \includegraphics[width=\linewidth]{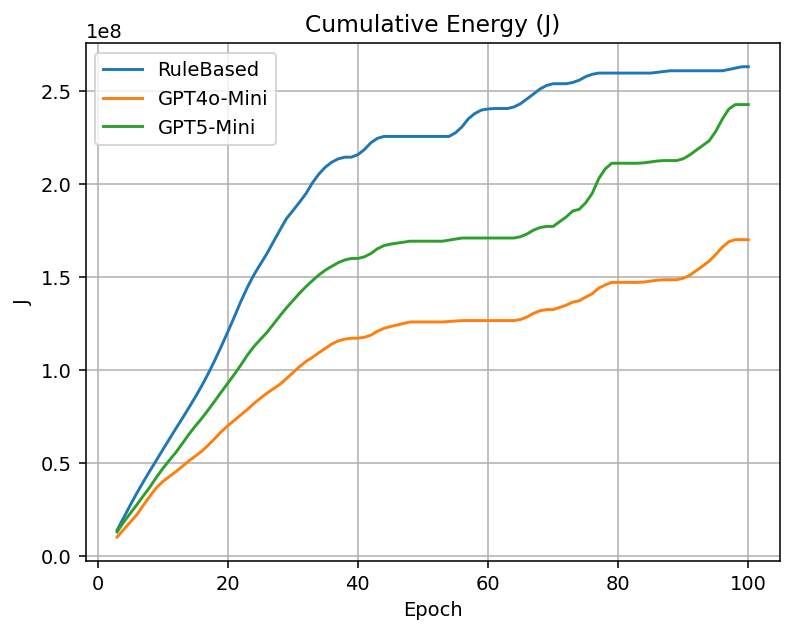}
    \caption{Energy}
  \end{subfigure}\hfill
  \begin{subfigure}{0.3\linewidth}
    \includegraphics[width=\linewidth]{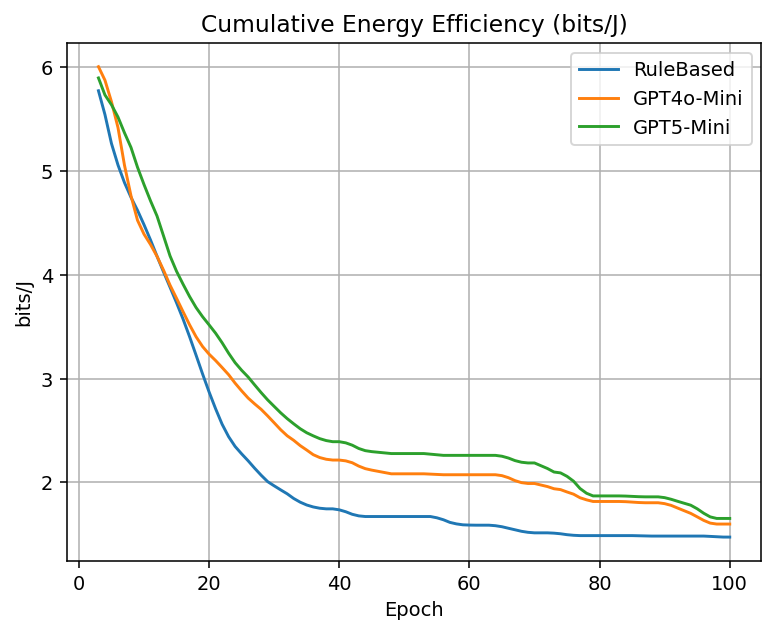}
    \caption{Energy Efficiency}
  \end{subfigure}
  \caption{Moderate load (40\,Mb/s) results.}
\end{figure*}

\subsection{High Offered Load (150\,Mb/s)}
With higher offered traffic the system remains service-limited for longer, and the differences between policies become more pronounced. The throughput curves indicate that \emph{GPT5-Mini} sustains the fastest cumulative growth and finishes with the highest total bits, while \emph{GPT4o-Mini} again trades some throughput for substantial energy savings. Total energy consumption is highest for the rule baseline across the entire horizon, reflecting aggressive duty usage that amplifies LBT loss; both LLM policies keep cumulative energy lower, and \emph{GPT5-Mini} delivers a favorable balance of bits and joules. The energy-efficiency curves mirror these trends: LLM-assisted control dominates the baseline throughout most of the run, with \emph{GPT5-Mini} providing the strongest long-horizon bits/J and \emph{GPT4o-Mini} offering the best energy containment when energy is the primary objective.

% \begin{figure}[t]
%   \centering
%   \includegraphics[width=0.5\linewidth]{150-throughput_all.png}
%   \vspace{-0.6em}
%   \caption{High load (150\,Mb/s): cumulative throughput over 100 epochs.}
%   \label{fig:150-throughput}
% \end{figure}

% \begin{figure}[t]
%   \centering
%   \includegraphics[width=0.5\linewidth]{150-energy_all.png}
%   \vspace{-0.6em}
%   \caption{High load (150\,Mb/s): cumulative energy over 100 epochs.}
%   \label{fig:150-energy}
% \end{figure}

% \begin{figure}[t]
%   \centering
%   \includegraphics[width=0.5\linewidth]{150-EE.png}
%   \vspace{-0.6em}
%   \caption{High load (150\,Mb/s): cumulative energy efficiency (bits/J).}
%   \label{fig:150-ee}
% \end{figure}

\begin{figure*}[ht]
  \centering
  \begin{subfigure}{0.3\textwidth}
    \includegraphics[width=\linewidth]{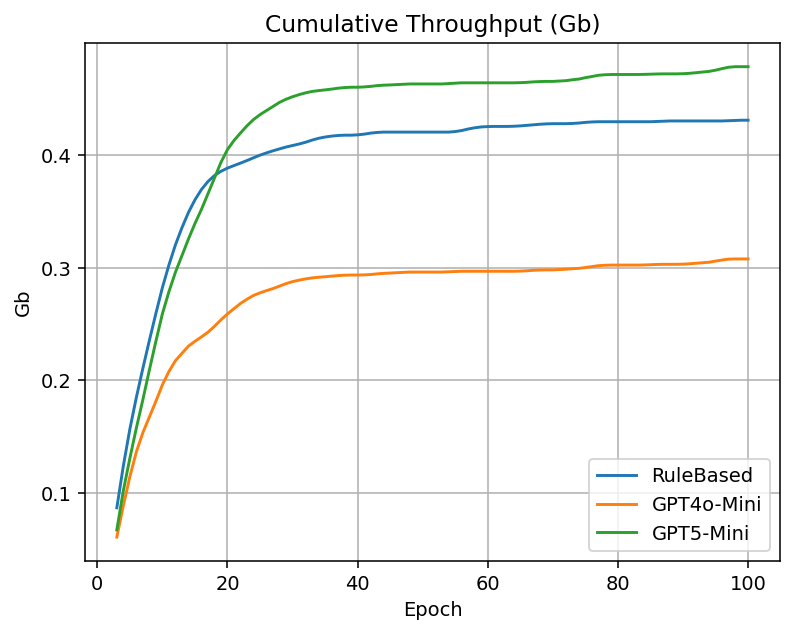}
    \caption{Cumulative Throughput (Gb)}
    \label{fig:150-throughput}
  \end{subfigure}\hfill
  \begin{subfigure}{0.3\textwidth}
    \includegraphics[width=\linewidth]{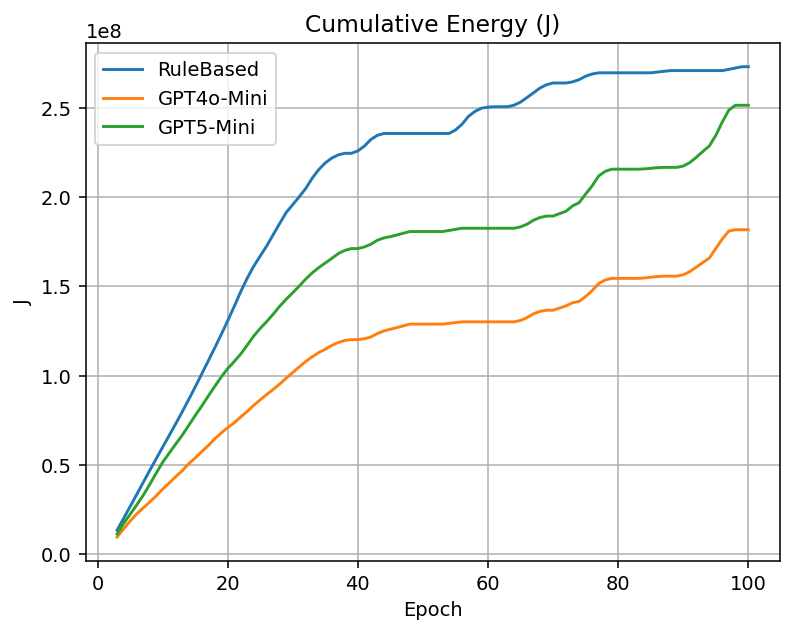}
    \caption{Cumulative Energy (J)}
    \label{fig:150-energy}
  \end{subfigure}\hfill
  \begin{subfigure}{0.3\textwidth}
    \includegraphics[width=\linewidth]{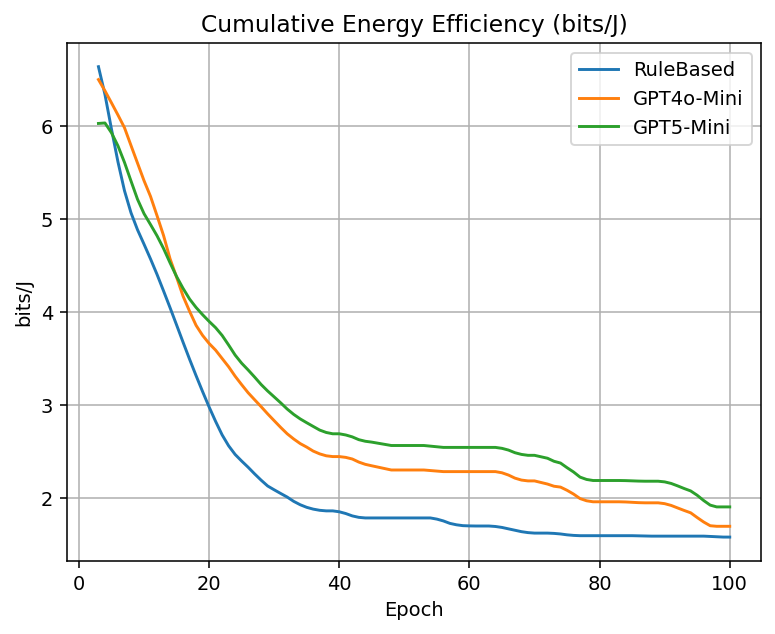}
    \caption{Cumulative Energy Efficiency (bits/J)}
    \label{fig:150-ee}
  \end{subfigure}
  \vspace{-0.5em}
  \caption{High offered load (150\,Mb/s) over 100 epochs. From left to right: cumulative throughput, cumulative energy, and cumulative energy efficiency.}
  \label{fig:150-row}
\end{figure*}

\subsection{Interpretation and Takeaways}
Across both load regimes, LLM-assisted policies consistently lower cumulative energy and improve bits/J relative to the rule baseline, while \emph{GPT5-Mini} achieves the best overall trade-off by pairing strong throughput with reduced energy. The qualitative shape of the curves is consistent with the agent’s design: LLM-proposed duty caps, combined with a fairness choice nearer to proportional fairness, keep the system away from congestion-dominated operating points where additional duty yields little goodput but incurs substantial energy. In the moderate-load setting, backlogs drain by mid-horizon and per-epoch throughput approaches zero for all methods; the cumulative differences observed up to that point therefore reflect more judicious early-phase decisions by the LLM policies. In the high-load setting, where the system remains busy, the LLM advantage persists throughout the horizon, indicating better long-run operating points under sustained traffic.

\section{Conclusion}
This paper introduced an LLM-assisted spectrum agent for Wi-Fi/NR-U coexistence in the 6\,GHz band. The core design cleanly separates high-level reasoning from verifiable execution: the policy layer---instantiated by a rule or by an LLM---chooses an $\alpha$-fairness regime, per-channel duty caps, and class weights from compact telemetry, while a deterministic optimizer enforces hard constraints and realizes post-LBT goodput and energy. This interface makes the role of the LLM transparent and auditable and guarantees safe control through clamping and rule fallback.
Across moderate and high offered loads, experiments show that LLM-assisted policies reduce cumulative energy and raise energy efficiency (bits/J) relative to a benevolent rule baseline, while maintaining competitive or superior cumulative throughput. The gains are most pronounced in the early and mid horizon, where headroom-aware caps and fairness choices near proportional fairness keep the system away from collision-dominated operating points. In a representative 100-epoch scenario, one LLM reduces total energy by more than a third, and another achieves the best overall trade-off with higher total bits and the highest bits/J among all methods.

\bibliographystyle{IEEEtran}
\bibliography{references}

\end{document}